\documentclass[12pt]{article}
\usepackage{scicite}
\usepackage{times}
\usepackage{graphics}
\usepackage{epsfig}
\usepackage{multirow}
\usepackage{ulem}
\usepackage{pdfpages}

\newenvironment{sciabstract}{%
\begin{quote} \bf}
{\end{quote}}

\newcounter{lastnote}

\title{Evidence for Two Distinct Populations \\ of Type Ia Supernovae}

\author{Xiaofeng~Wang$^{1,2,\ast}$, Lifan~Wang$^{2,3}$, Alexei V. Filippenko$^{4}$, \\
 Tianmeng~Zhang$^{5}$, Xulin~Zhao$^{1}$ \\
\scriptsize{$^{1}$Department of Physics, Tsinghua University, Beijing, 100084, China} \\
\scriptsize{$^{2}$Mitchell Institute for Fundamental Physics and Astronomy,
Texas A\&M University, College Station, TX 77843, USA} \\
\scriptsize{$^{3}$Purple Mountain Observatory, Nanjing, 201008, Jiangsu, China} \\
\scriptsize{$^{4}$Department of Astronomy, University of California, Berkeley, CA 94720-3411, USA} \\
\scriptsize{$^{5}$National Astronomical Observatory of China, Chinese Academy of Sciences, Beijing, 100012, China} \\
\\
\scriptsize{$^\ast$To whom correspondence should be addressed; E-mail: wang\_xf@mail.tsinghua.edu.cn}}

\date{}

\begin{document}


\baselineskip24pt

\maketitle

\begin{sciabstract}
Type Ia supernovae (SNe~Ia) have been used as excellent standardizable
candles for measuring cosmic expansion, but their progenitors are
still elusive. Here we report that the spectral diversity of SNe~Ia is
tied to their birthplace environments. We find that those with
high-velocity ejecta are substantially more concentrated in the inner
and brighter regions of their host galaxies than are normal-velocity
SNe~Ia. Furthermore, the former tend to inhabit larger and more-luminous
hosts. These results suggest that high-velocity SNe~Ia likely originate
from relatively younger and more metal-rich progenitors than normal-velocity
SNe~Ia, and are restricted to galaxies with substantial chemical evolution.

\end{sciabstract}


Type Ia supernovae (SNe~Ia) are among the most energetic and
relatively uniform stellar explosions in the Universe, and were used
to discover its accelerating expansion ({\it 1, 2}). They are thought
to originate from a thermonuclear explosion of an accreting
carbon-oxygen white dwarf (WD) near the Chandrasekhar mass limit
($M_{\rm Ch} \approx 1.4$\,M$_\odot$) in a close binary system ({\it
  3, 4}). Two competing scenarios have been proposed for the
progenitor systems: single-degenerate (SD) ({\it 5, 6}) and
double-degenerate (DD) models ({\it 4, 7}). In the former, the
mass-donating star could be a main-sequence (MS)/subgiant star ({\it 8}),
a red-giant star (RG; {\it 9}), or even a helium star ({\it 10, 11}), while it
is another WD in the latter scenario ({\it 4, 7}). Recent results suggest that
both scenarios are possible ({\it 12-18}).

There is increasing evidence for spectral diversity among SNe~Ia. Of
particular interest are those showing higher expansion velocities as
inferred from the blueshifted Si~II 615\,nm feature in optical spectra
({\it 19}). These fast-expanding SNe~Ia also generally exhibit a steep Si~II
temporal velocity gradient ({\it 20}). This spectral difference in
velocity or velocity evolution of the ejecta has been proposed to be a
geometric effect of an asymmetric explosion ({\it 21, 22}). Given a
common origin for SNe~Ia having different ejecta velocities, they
should be found in similar stellar environments. This can be tested by
examining SN positions in their hosts, the surface brightness at these
locations, and the properties of their hosts.

We conducted such an analysis with a well-defined SN sample having 188
SNe~Ia [see supporting online material (SOM) text S1] from the Lick
Observatory Supernova Search (LOSS; {\it 23}). The SN~Ia sample
consists of 123 ``Branch-normal" (spectroscopically normal) objects
({\it 24}), 30 peculiar ones of the SN 1991bg variety ({\it 25}), 13
peculiar ones like SN 1991T ({\it 26, 27}), and 7 peculiar ones like
SN 2002cx ({\it 28}), with respective fractions of 65.4\%, 16.0\%,
6.9\%, and 3.7\% (Table S1). There are 15 SNe~Ia (8.0\% of all) that
cannot be subclassified due to an absence of early-time spectra. We
concentrate on the Branch-normal SNe~Ia which are thought to be
relatively uniform. We obtained the main parameters of the host galaxies
from two large online astronomical databases: the NASA/IPAC Extragalactic
Database (NED; {\it 29}) and HyperLeda ({\it 30}).

The location of a SN in its host galaxy can be estimated by the radial
distance of the SN from the nucleus ($R_{\rm SN}$). Assuming that the
galaxies are circular disks and only appear to have different major
and minor axes due to their inclination, $R_{\rm SN}$ can be
calculated if we know the position angle and the axial ratio of each
galaxy. The radius of the galaxy ($R_{\rm gal}$) is simply the
semimajor axis at the 25.0 $B$-mag arcsec$^{-2}$ isophote. The
ratio $R_{\rm SN}/R_{\rm gal}$ is then the fractional radial distance
of the SN. For SNe~Ia in elliptical galaxies, no tilt correction is applied
because these galaxies can be regarded as spheroids. The typical host galaxy
of our sample (Fig. S1) has a major axis of about $1.3'$--$1.4'$ and can be
measured with a precision of $\sim 0.1'$. This results in a typical
uncertainty of $\sim 0.05$ in the determination of $R_{\rm SN}/R_{\rm gal}$.

We measured the velocity of the Si~II 615\,nm line for 165 SNe~Ia (out
of 188) by using the published spectral datasets ({\it 31--33}). We
normalized this velocity to the maximum-light value with a series of
templates of Si~II velocity evolution established from well-observed
SNe~Ia, with a typical uncertainty of 300--400\,km\,s$^{-1}$
(SOM text S1 and Fig. S2).

It is clear that Branch-normal SNe~Ia with $v_{\rm Si~II} <$ 12,000\,km\,s$^{-1}$
(the normal-velocity group, NV) span a wide radial distribution, occurring
at places from the innermost region to about 2--3 times the optical radius of
the entire galaxy. In contrast, those with $v_{\rm Si~II} \geq$ 12,000\,km\,s$^{-1}$
(the high-velocity group, HV) are rarely found at large galactic radii (Fig.1b).
For example, only 3 out of the 40 HV SNe~Ia are detected in regions with
$R_{\rm SN}/R_{\rm gal} > 0.7$ (2 of which are in elliptical galaxies),
whereas $14 \pm 2$ would have been expected at the detection rates of
the NV SNe~Ia (which are about $34 \pm 5$\% at $R_{\rm SN}/R_{\rm gal}
> 0.7$). Such a difference has a statistical significance of about
5$\sigma$, highlighting the paucity of HV SNe~Ia in outskirts of galaxies.
Binning the data in velocity space with an interval of about 2000\,km\,s$^{-1}$
further shows a correlation between the ejecta velocity of SNe~Ia and the
locations in their host galaxies (see gray quadrangles in Fig. 1a).

To better understand such a birthplace vs. ejecta-velocity relation
for SNe~Ia, it is important to know how the Si~II velocity itself is
distributed. Fig. 1c shows that most of the sample of Branch-normal
SNe~Ia clusters at velocities between 10,000 and 12,000\,km\,s$^{-1}$,
with a tail extending up to $\sim$ 16,000\,km\,s$^{-1}$. Such a velocity
distribution can be fit by a double-Gaussian model. One component, with a
stronger and narrower peak at 10,800\,km\,s$^{-1}$, is responsible for the
NV group, whereas the other component, with a weaker and broader peak at
13,000\,km\,s$^{-1}$, accounts for the HV group. Adopting a velocity cut of
$v_{\rm Si~II} =$ 12,000\,km\, s$^{-1}$ to divide these two groups puts 40
SNe~Ia in the HV group and 83 in the NV group. Accordingly, we estimate
the fraction of the HV population to be 1/3 of the Branch-normal sample and
1/5--1/4 of the entire SN~Ia sample. Note that the Si~II velocity distinction
is not sharp between these two groups, so blending could occur to some extent.
Nevertheless, this blending is small at larger velocities, and it will not
affect the result that the SNe~Ia with higher velocities tend to occur nearer
the galaxy centers.

HV SNe~Ia show the highest central concentration among the samples, with
90\% occurring in regions within $R_{\rm SN}/R_{\rm gal} \leq 0.7$; this
fraction is 66\%, 77\%, and 89\% for NV SNe~Ia, SNe~II, and SNe~Ibc,
respectively (Fig.2). The contrast is even more significant when examining
the distribution toward the galaxy center. We caution, however, that a
higher fraction of CC~SNe may be missing in the central regions of galaxies
during LOSS because of their lower luminosity relative to SNe~Ia.

A Kolmogoroff-Smirnoff (K-S) test finds a probability of 0.5\% that HV
and NV SN~Ia groups have a similar radial distribution in galaxies.
This probability further decreases to 0.1\% if one increases the velocity
cut dividing these two groups from 12,000\,km\,s$^{-1}$ to 13,000 \,km\,s$^{-1}$.
Possible selection effects in the radial distribution have been explored
and none can account for such a significant discrepancy between HV and
NV SN~Ia groups (SOM text S2 and Figs. S3). Thus HV SNe~Ia have higher
metallicities and their progenitors are therefore less likely to be
from the halo population that consists of old, metal-poor stars located
far away from the galactic center ({\it 34}).

Because the radial distances give only a rough estimate of the
properties of SN progenitors, more sophisticated methods are necessary
to provide additional constraints. A simple statistic of the
``fractional flux" allows a measurement of how SNe are distributed
within their hosts ({\it 35}). This can be achieved by measuring in
the host-galaxy images the fraction of total galaxy light contained in
pixels fainter than or equal to the light in the pixel at the location
of the SN. The ``fractional flux'' was obtained with the SDSS
$u'g'r'$-band images ({\it 36}) for 64 SNe~Ia (39 NV + 25 HV), 102
SNe~II, and 39 SNe~Ibc of the entire SN sample (SOM text S3).

The surface brightness of SN~II locations has an approximately linear
distribution, roughly tracing the distribution of the light in their
hosts (Fig. 3). SNe~Ibc seem to be more concentrated in brighter regions
of the host galaxies, consistent with the knowledge that they arise in
larger star-forming regions that produce more-massive stars. Of the SN~Ia
sample, the locations of the HV group and the NV group track their
hosts' light differently, with a very low probability ($P = 0.04$ in
$u'$, 0.07 in $g'$, and 0.08 in $r'$) that they come from the same stellar
populations. Figure 3 shows that the NV SN~Ia locations are apparently
fainter than those of CC~SNe; the HV SN~Ia distribution, on the other
hand, is similar to the CC~SN distribution.

The light radii and luminosities of galaxies in the HV sample and the NV
sample also differ significantly (Fig. 4). A two-dimensional K-S test gives a
probability of 0.5\% that they come from the same population. The mean-light
radius of the hosts estimated for these two groups is $20.67 \pm 0.83$\,kpc
for HV and $16.61 \pm 0.67$\,kpc for NV. The mean absolute $K$-band magnitudes
are $-24.53 \pm 0.13$\,mag (HV) and $-24.17 \pm 0.10$\,mag (NV), with the
HV hosts being, on average, brighter by about 40\%. In general, the HV
SNe~Ia tend to occur in larger and more luminous hosts, and the fraction
found in galaxies with $R_{\rm gal} < 15$\,kpc is very low, 15\% (versus
46\% for the NV counterparts). This difference is not due to an observational
bias because the host galaxies of these two groups do not show significantly
different distributions in either morphologies or redshifts (see SOM text
S4 and Figs. S4 and S5).

Our analysis thus reveals significant differences in the progenitor
environments of SNe~Ia having different Si~II velocities. The HV
SNe~Ia are much more concentrated in regions close to the galaxy center
and in bright regions of their host galaxies, and they also tend to
reside in larger and more luminous hosts relative to the NV group. It
is generally accepted that all galaxies (on average) have metallicities
that systematically decrease outward from their galactic centers, and that
their global metallicities increase with galaxy size ({\it 39}). Higher
metallicity is therefore expected for the HV SN~Ia progenitor population.
Meanwhile, the fact that the surface brightness at HV SN~Ia locations roughly
traces the light of their host galaxies, as with CC~SNe, suggests that the
progenitor populations are relatively young. Thus, the HV SN~Ia population may
have a younger and more metal-rich progenitor system than the NV SN~Ia population,
and a larger initial MS mass of the exploding WD may be expected for the
former due to the shorter evolutionary time needed before explosion.

Calculations of stellar evolution show that both stellar mass and
metallicity have significant effects on the nature of C-O WDs that may
become progenitors of SNe~Ia. The maximum MS mass to form C-O WDs is found
to increase significantly toward higher metallicity ({\it 40}); stars
with MS masses up to 8--9\,M$_{\odot}$ produce massive WDs
($\sim 1.1$\,M$_{\odot}$) that can reach 1.4\,M$_{\odot}$ in a shorter
time via mass transfer from a companion star. Thus, the HV SN~Ia
group might represent the young-population SNe~Ia, corresponding to the
``prompt'' component with short delay times ({\it 41, 42}), while the
NV group may belong to the ``older'' component with long delay times. We
note that 5 out of the 40 (12.5\%) SNe~Ia in the HV sample are in elliptical
galaxies that are luminous (see Table S1 and Fig. S4) and massive. However,
among these, some (e.g., SN 2002dj and SN 2000B) show dust structure and
molecular gas {\it 43, 44}), suggestive of recent or ongoing star
formation in them; they do not contradict the conclusion that HV SNe~Ia
likely arise from young stellar populations.

Having young and metal-rich progenitors may explain the numerous recent
detections of circumstellar medium (CSM) signatures among HV SNe~Ia.
Time-variable absorption features of the Na~I doublet
(D1 589.6 nm and D2 589.0 nm), likely suggestive
of changes in CSM ionization due to a variable SN radiation field, have been
reported for a few SNe~Ia such as SNe 2006X, 2007le, and 1999cl
({\it 12, 45, 46}). A common feature of these SNe~Ia is that they belong
to the HV subclass (see Table S1). Additionally, the velocity structure of
the line-of-sight Na~I lines provides another possible diagnostic of CSM
around SNe. A trend of blueshifts was found among SNe~Ia based on a
study of Na~I absorption lines of a larger SN sample ({\it 13}).
It was noticed, however, that the SNe~Ia with a blueshifted absorption feature
generally have higher Si~II velocities (47). This evidence is
consistent with their systematically redder color around maximum light
(perhaps due to additional CSM absorption) relative to the NV population ({\it 19}),
and it can be understood in terms of an empirical metallicity dependence of mass
outflow ({\it 48}). At higher metallicity, stars lose more mass and produce
stronger outflows than their lower-metallicity counterparts. Thus, detection of
abundant outflows in the vicinity of the HV SNe~Ia is not unexpected.

The higher Si~II velocity seen in HV SNe~Ia may be related in part to
an increase in stellar metallicity. As the metallicity increases in
the C+O layer of the exploding WDs, the line-forming region moves
toward shallower parts of the atmosphere because of an increased line
opacity, leading to larger line velocities ({\it 49}). In addition,
the high-velocity Si~II layers are perhaps formed due to the density
increase caused by interaction between the SN ejecta and the material
around the exploding WD, which could be an accretion disk or a filled
Roche lobe ({\it 50}). The angular variations in observing such an
interacting system may explain the variation of the polarization
across the Si~II line ({\it 51}) and its correlation with the ejecta
velocity for HV SNe~Ia ({\it 22}). This is consistent with the result
that the observed differences in some SNe~Ia might be due to a projection
effect ({\it 21}). However, these data and analyses are insensitive to a
separate population of NV SNe~Ia that is intrinsically different from HV SNe~Ia.

The dependence of SN~Ia ejecta velocity on progenitor environment could be
relevant when using SNe~Ia as cosmological yardsticks, because the HV and NV
populations have different colors around maximum light ({\it 19}) and their
ratio may change with redshift. The observed relative fraction of the HV
and NV population might become smaller at great distances due to a decrease
in the HV SN~Ia rate in low-metallicity environments and the increased difficulty of
spectroscopically classifying SNe in the central regions of distant galaxies.

\newpage
\subsection*{References and Notes}
\begin{itemize}
\item[1.] A. G. Riess, et al., {\it Astron. J.} {\bf116}, 1009 (1998).
\item[2.] S. Perlmutter, et al., {\it Astrophys. J.} {\bf 517}, 565 (1999).
\item[3.] K. Nomoto, {\it Astrophys. J.} {\bf 253}, 798 (1982).
\item[4.] I. Iben, Jr. \& A. V. Tutukov, {\it Astrophys. J. Suppl. Ser.} {\bf 54}, 335 (1984).
\item[5.] K. Nomoto, K. Iwamoto, \& N. Kishimoto, {\it Science} {\bf 276}, 1378 (1997).
\item[6.] P. Podsiadlowski, P. Mazzali, P. Lesaffre, Z. Han, \& F. F{\"o}rster, {\it N. Astro. Rev.} {\bf 52}, 381 (2008).
\item[7.] R. F. Webbink, {\it Astrophys. J.} {\bf 277}, 355 (1984).
\item[8.] E. P. J. van den Heuvel, D. Bhattacharya, K. Nomoto, \& S. A. Rappaport, {\it Astron. Astrophys.} {\bf 262}, 97 (1992).
\item[9.] U. Munari \& A. Renzini, {\it Astrophys. J.} {\bf 397}, L87 (1992).
\item[10.] B. Wang \& Z. W. Han, {\it Astron. Astrophys.} {\bf 515}, A88 (2010).
\item[11.] S.-C. Yoon \& N. Langer, {\it Astron. Astrophys.} {\bf 412}, L53 (2003).
\item[12.] F. Patat, et al., {\it Science} {\bf 317}, 924 (2007).
\item[13.] A. Sternberg, A. Gal-Yam, \& J. D. Simon, {\it Science} {\bf 333}, 856 (2011).
\item[14.] P. Nugent, et al., {\it Nature} {\bf 480}, 344 (2011).
\item[15.] W. Li, et al., {\it Nature} {\bf 480}, 348 (2011).
\item[16.] J. Bloom, et al., {\it Astrophys. J.} {\bf 744}, L17 (2012).
\item[17.] P. Brown, et al., {\it Astrophys. J.} {\bf 753}, 22 (2012).
\item[18.] B. Schaefer \& A. Pagnotta, {\it Nature} {\bf 481}, 164 (2012).
\item[19.] X. Wang, et al., {\it Astrophys. J.} {\bf 699}, L139 (2009).
\item[20.] S. Benetti, et al., {\it Astrophys. J.} {\bf 623}, 1011 (2005).
\item[21.] K. Maeda, et al., {\it Nature} {\bf 466}, 82 (2010).
\item[22.] J. R. Maund, et al., {\it Astrophys. J.} {\bf 725}, L167 (2011).
\item[23.] J. Leaman, W. Li, R. Chornock, \& A. V. Filippenko, {\it Mon. Not. R. Astron. Soc.} {\bf 412}, 1419 (2011).
\item[24.] D. Branch, A. Fisher, \& P. Nugent, {\it Astron. J.} {\bf 106}, 2383 (1993).
\item[25.] A. V. Filippenko, et al., {\it Astron. J.} {\bf 104}, 1543 (1992).
\item[26.] A. V. Filippenko, et al., {\it Astrophys. J.} {\bf 384}, 15 (1992).
\item[27.] M. M. Phillips, et al., {\it Astron. J.} {\bf 103}, 1632 (1992).
\item[28.] W. Li, et al., {\it Publ. Astron. Soc. Pac.} {\bf 115}, 453 (2003).
\item[29.] http://ned.ipac.caltech.edu/ .
\item[30.] http://leda.univ-lyon1.fr/ .
\item[31.] S. Blondin, et al., {\it Astron. J.} {\bf 143}, 126 (2012).
\item[32.] J. M. Silverman, et al., {\it Mon. Not. R. Astron. Soc.} {\bf 425}, 1789 (2012).
\item[33.] T. Matheson, et al., {\it Astrophys. J.} {\bf 135}, 1598 (2008).
\item[34.] K. M. Gilbert, et al. {\it Astrophys. J.} {\bf 652}, 1181 (2006).
\item[35.] A. S. Fruchter, et al., {\it Nature} {\bf 441}, 463 (2006).
\item[36.] D. G. York, et al., {\it Astron. J.}, {\bf 120}, 1579 (2000).
\item[37.] D. N. Spergel, et al., {\it Astrophys. J. Suppl. Ser.} {\bf 170}, 377 (2007).
\item[38.] D. J. Schlegel, D. P. Finkbeiner, \& M. Davis, {\it Astrophys. J.} {\bf 500}, 525 (1998).
\item[39.] R. B. C. Henry \& G. Worthey, {\it Publ. Astron. Soc. Pac.} {\bf 111}, 919 (1999).
\item[40.] H. Umeda, et al., {\it Astrophys. J.} {\bf 513}, 861 (1999).
\item[41.] E. Scannapieco \& L. Bildsten, {\it Astrophys. J.} {\bf 629}, L85 (2005).
\item[42.] F. Mannucci, M. Della Valle, \& N. Panagia, {\it Mon. Not. R. Astron. Soc.} {\bf 370}, 773 (2006).
\item[43.] P. Goudfrooij, L. Hansen, H. E. Jorgensen, et al. {Astron. Astrophys. Supp. Ser.} {\bf 105} 341 (1994).
\item[44.] L. M. Young, et al., {\it Mon. Not. R. Astron. Soc.} {\bf 414}, 940 (2011).
\item[45.] J. D. Simon, et al., Astrophys. J. 702, 1157 (2009).
\item[46.] S. Blondin, et al., Astrophys. J. 693, 207 (2009).
\item[47.] R. J. Foley, et al., {\it Astrophys. J.} {\bf 752}, 101 (2012).
\item[48.] P. R. Wood, {\it ASP Conf. Ser.} {\bf 404}, 255 (2009).
\item[49.] E. J. Lentz, et al., {\it Astrophys. J.} {\bf 530}, 966 (2000).
\item[50.] C. Gerardy, et al., {\it Astrophys. J.} {\bf 607}, 391 (2004).
\item[51.] L. Wang, D. Baade, \& F. Patat, {\it Science} {\bf 315}, 212 (2007).
\end{itemize}

\newpage
\begin{figure}
\vspace{-2.5cm}
\includegraphics[width=16.0cm]{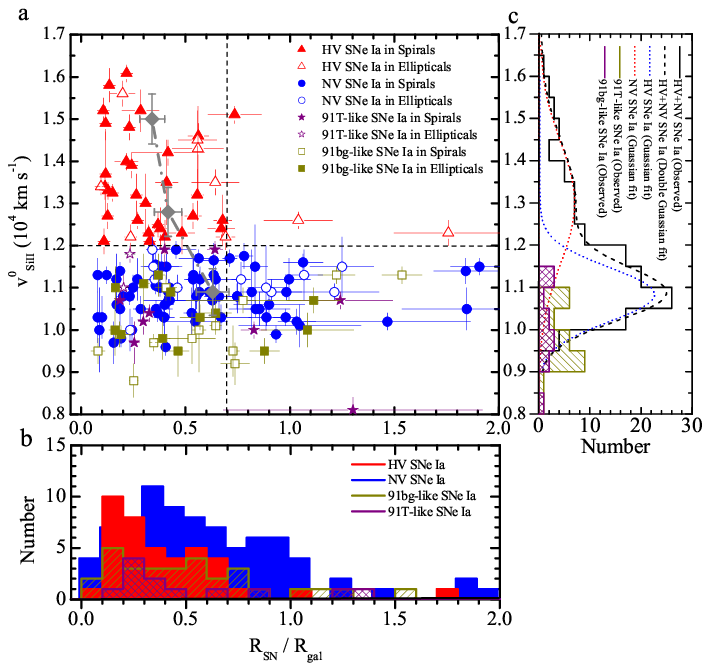} \\
\vspace{-0.0cm}
\noindent {\bf Fig. 1:} Relations between the Si~II velocity of SNe~Ia
and the birth location in their host galaxies. {\bf a,} The Si~II
velocity obtained around $B$-band maximum light ($v^{0}_{\rm Si~II}$,
ordinate) as compared to the fractional radial distance in the host
galaxy ($R_{\rm SN}/R_{\rm gal}$, abscissa) for 165 SNe~Ia. The
Branch-normal SNe Ia with $v^{0}_{\rm Si~II}$ $\geq$ 12,000 km
s$^{-1}$ (HV group), those with $v^{0}_{\rm Si~II}$ $<$ 12,000 km
s$^{-1}$ (NV group), SN 1991T-like SNe, and SN 1991bg-like SNe are
shown by red triangles, blue dots, purple stars, and dark-gray squares,
respectively. The SNe~Ia in spiral and elliptical/lenticular galaxies
are represented with filled and open symbols, respectively. The gray
quadrangles show the radial distances averaged in binned velocity
space, which are $0.63 \pm 0.08$ in 9,000--12,000\,km\,s$^{-1}$, $0.42
\pm 0.06$ in 12,000--14,000\,km\,s$^{-1}$, and $0.34 \pm 0.06$ in
$\sim$ 14,000--16,000\,km\,s$^{-1}$, respectively.  The horizontal and
vertical dashed lines mark the place with $v_{\rm Si~II}$ =
12,000\,km\,s$^{-1}$ and with $R_{\rm SN}/R_{\rm gal} = 0.7$. {\bf b},
The number distribution of the fractional radial distance. The red and
blue areas are for the HV and NV groups of SNe~Ia. The purple and dark-gray
areas are for the SN 1991T-like and SN 1991bg-like SNe~Ia. {\bf c}, The
number distribution of near-maximum-light Si~II velocity. A
double Gaussian function is used to fit the distribution of 123
Branch-normal (HV + NV) SNe~Ia. Red and blue curves are for the
high-velocity and normal-velocity components, with respective peaks
centered at 13,000\,km\,s$^{-1}$ and 10,800\,km\,s$^{-1}$. The black
curve represents the combined result of these two components.
\end{figure}

\newpage
\begin{figure}
\vspace{0 cm}
\includegraphics[width=16.0cm]{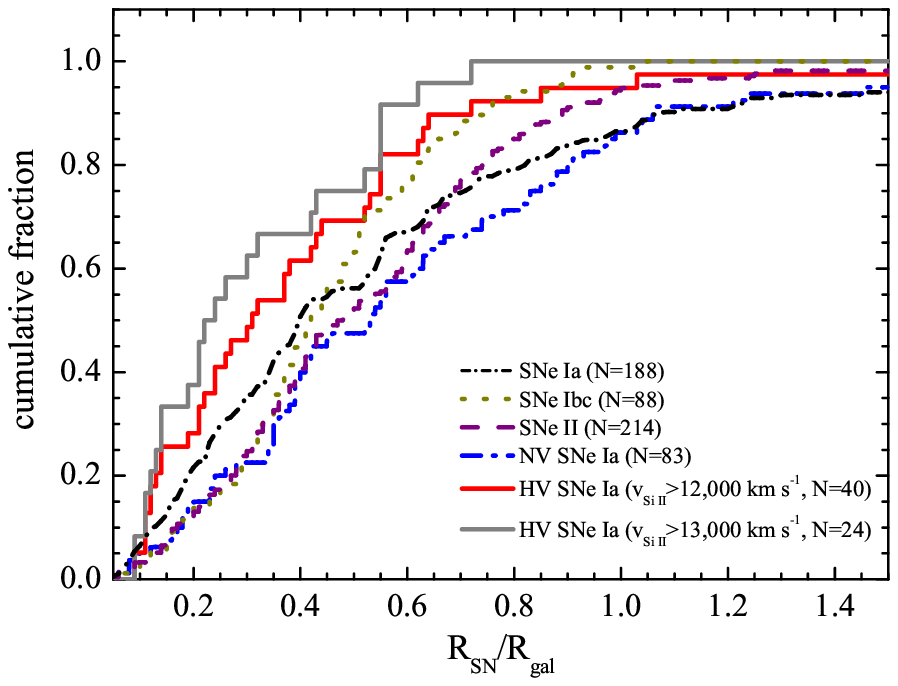} \\
\noindent {\bf Fig. 2:}  A plot of the cumulative fraction of our SN
samples (HV SNe~Ia, NV SNe~Ia, SNe~II, and SNe~Ibc). The gray solid curve
represents the distribution of SNe~Ia with $v_{\rm Si~II} >$ 13,000\,km\,s$^{-1}$.
\end{figure}

\newpage
\begin{figure}
\vspace{0 cm}
\includegraphics[width=16.0cm]{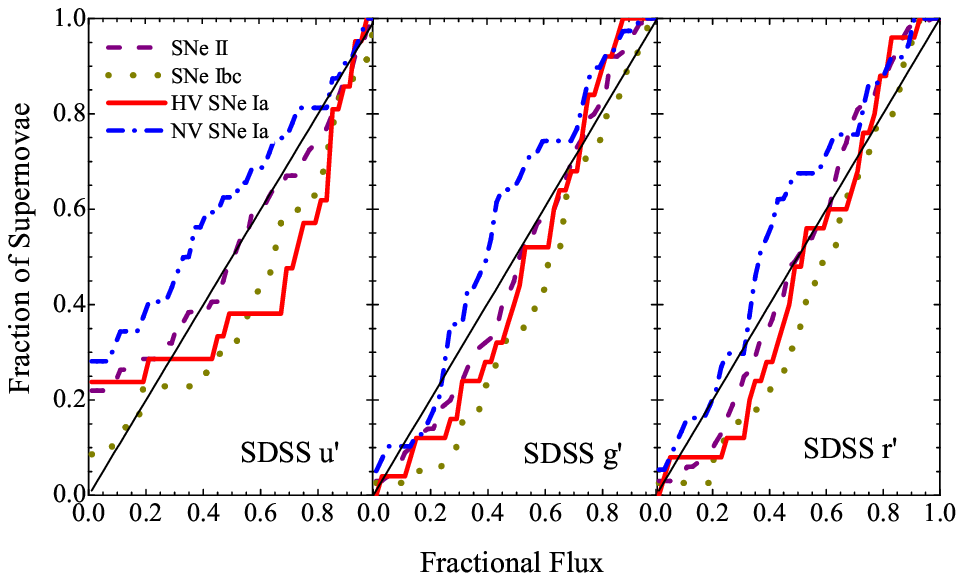} \\
\noindent {\bf Fig. 3:} Histogram distribution of the fractional flux
of the host-galaxy light at the location of SN explosions in the $u'$
(left panel), $g'$ (middle panel), and $r'$ (right panel) bands.
The diagonal black lines represent the case that the SN progenitors
follow exactly the distribution of galaxy light. The HV SN~Ia population 
differs significantly from the NV SN~Ia population at the SN location 
(with a respective probability of 3.8\% in $u'$, 6.8\% in $g'$, and 
8.3\% in $r'$ that the SNe come from the same radial distribution), 
but seems to have a distribution similar to
those of SNe~II and SNe~Ibc..
\end{figure}

\newpage
\begin{figure}
\vspace{0 cm}
\includegraphics[width=16.0cm]{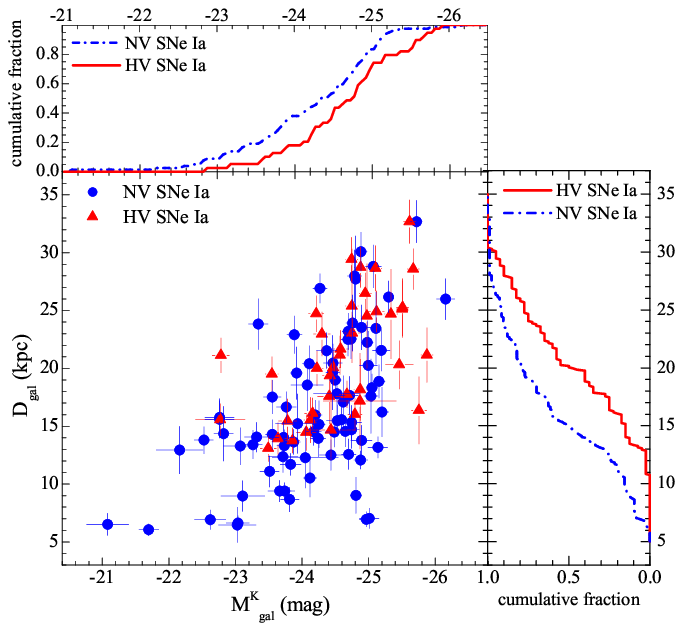} \\
\noindent {\bf Fig. 4:} A comparison of the physical sizes ($B$-band light radius at
25\,mag\,arcsec$^{-2}$ isophote) and absolute $K$-band magnitude
distributions of the SN hosts. These two parameters have been derived
by adopting $\Omega_{\rm m} = 0.27$, $\Omega_{\Lambda} = 0.73$, and
H$_{0}$ = 73\,km\,s$^{-1}$\,Mpc$^{-1}$ ({\it 37}). Foreground Galactic
absorption corrections ({\it 38}) have been applied to the absolute
magnitudes. In the main panel, the high-velocity SN~Ia hosts are represented
as red triangles, and the normal-velocity SN~Ia hosts as blue dots. The
absolute magnitudes of the hosts are shown on the abscissa, and the lengths
of the semimajor axes of the hosts on the ordinate. The plot is then
projected onto the two side panels where a histogram is displayed for
each host population in each of the dimensions, absolute magnitude
and semimajor axis.
\end{figure}

\subsection*{Acknowledgments.}
We thank I. Dominguez for helpful discussions and C. WU for assistance with the
SDSS images. This work is supported by the National Natural Science Foundation
of China (NSFC grants 11073013, 11178003), the Major State Basic Research
Development Program (2009CB824800), and the Foundation of Tsinghua University
(2011Z02170). The work of L.W. is supported by NSF grant AST-0708873. A.V.F.
is grateful for financial assistance from NSF grant AST-1211916, the TABASGO
Foundation, and the Christopher R. Redlich Fund. We dedicate this paper to the
memory of our dear friend and colleague, Weidong Li, whose unfailing dedication
to the Lick Observatory Supernova Search made this work possible; his premature,
tragic passing has deeply saddened us.

\newpage
\subsection*{Supporting Online Material}

\vskip 1.0 cm
{\bf S1. The Supernova Sample}
\vskip 0.5 cm

The SN sample used in our study is the "season-optimal" subsample from the Lick Observatory Supernova Search (LOSS), found over the interval 1998?2008 (23). This subsample contains 499 SNe that were discovered only during (not prior to) the active monitoring period of their host galaxies, with those in small early-type and edge-on spiral galaxies further excluded (see Table 4 of Ref. 23).  Fig. S1 shows the typical host galaxies of the "season-optimal" subsample used in our study. Among the 499 SNe, 188 are SNe Ia (37.7\%), 214 are SNe II (43.1\%), 88 are SNe Ibc (17.6\%), and 9 have no known spectroscopic classification (1.8\%). These fractions of different SN types are fully consistent with those from the full "optimal" sample (726 in total, of which 284 are SNe Ia) used for the final rate calculations for nearby SNe (52), arguing against a possible bias in the sample selection. The main advantage of adopting the smaller "season-optimal" subsample is that the discovered SNe are generally at a relatively young phase, allowing accurate, timely spectroscopic classifications of them. This is particularly important for the identification of spectral diversity among SNe  Ia.

\begin{figure}
\includegraphics[angle=270, width=155 mm]{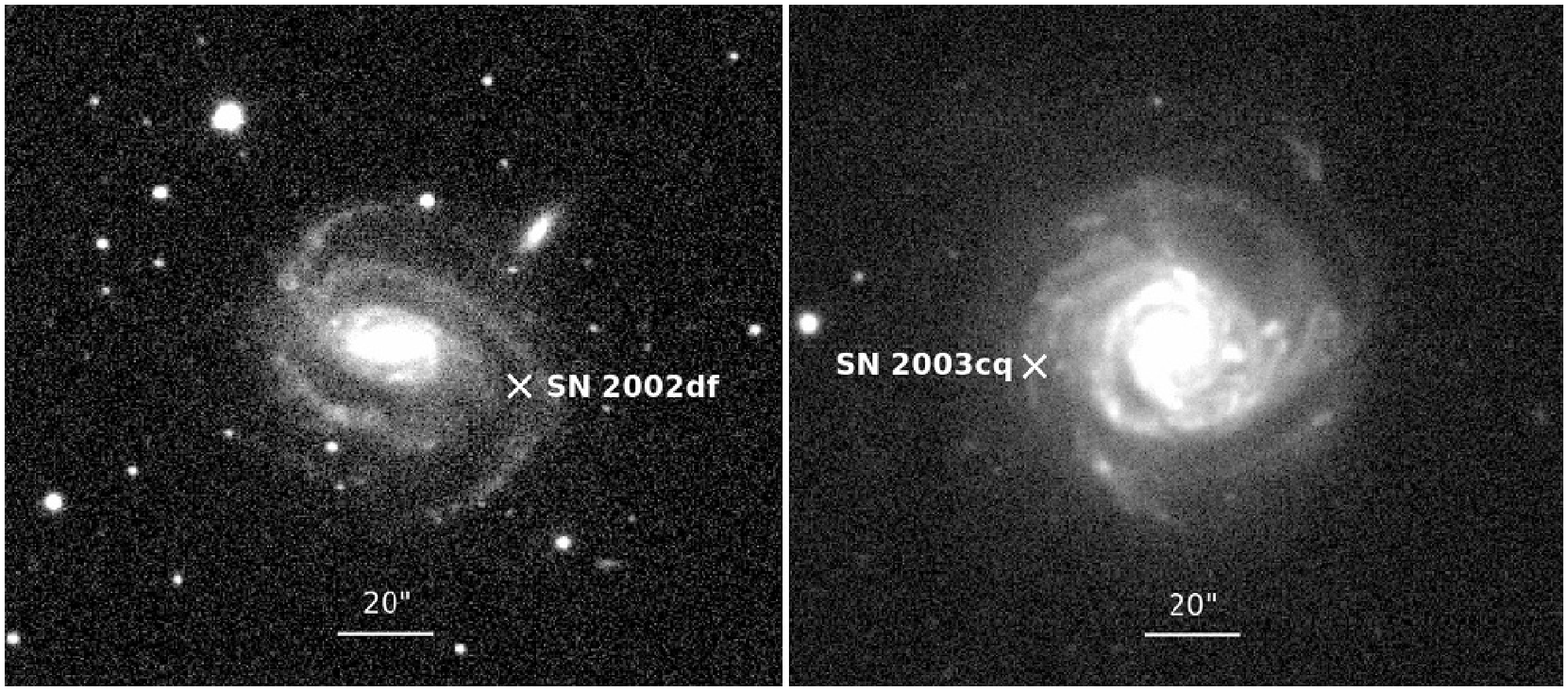} \\
\noindent
{\bf Fig. S1:}
Typical galaxies in the sample. Left: SDSS g'$-$band image of the host galaxy MCG-1-53-6 of SN 2002df, 
a member of the NV population of SNe Ia. Right: SDSS g'$-$band image of the host galaxy NGC 3978 of 
SN 2003cq, a member of the HV population of SNe Ia. Both images have a size of 2.8'$\times$2.5'.
\end{figure}

The ejecta velocity can be measured from the absorption minimum of the blueshifted Si~II 635.5 nm line, which is one of the strongest features in early-time optical/near-infrared spectra of SNe Ia and suffers from little blending with other features.  In order to measure an accurate expansion velocity from the Si~II lines, we smoothed the observed spectrum with a locally weighted scatter-plot smoothing method ({\it 53}) to avoid taking a dip of a noise spike in the data as a local absorption minimum. As shown in the inset of Fig. S2, the wavelength of the minimum in the smoothed spectrum ( min) is used to calculate the expansion velocity using the special-relativistic formula. We further compare these velocities obtained within about one week from maximum light to templates of Si~II velocity evolution to determine the corresponding values at maximum light, v$^{0}_{Si~II}$ (see Fig. S2). These comparison templates are established from some well-observed objects with published spectral data (19, 31-33, and references therein). The values of v$^{0}_{Si~II}$ used in this study are listed in Table S1, with a typical uncertainty of about 300-400 km s$^{-1}$, depending on the number, epoch, and quality of the available spectra.

\begin{figure}
\includegraphics[width=16.0cm]{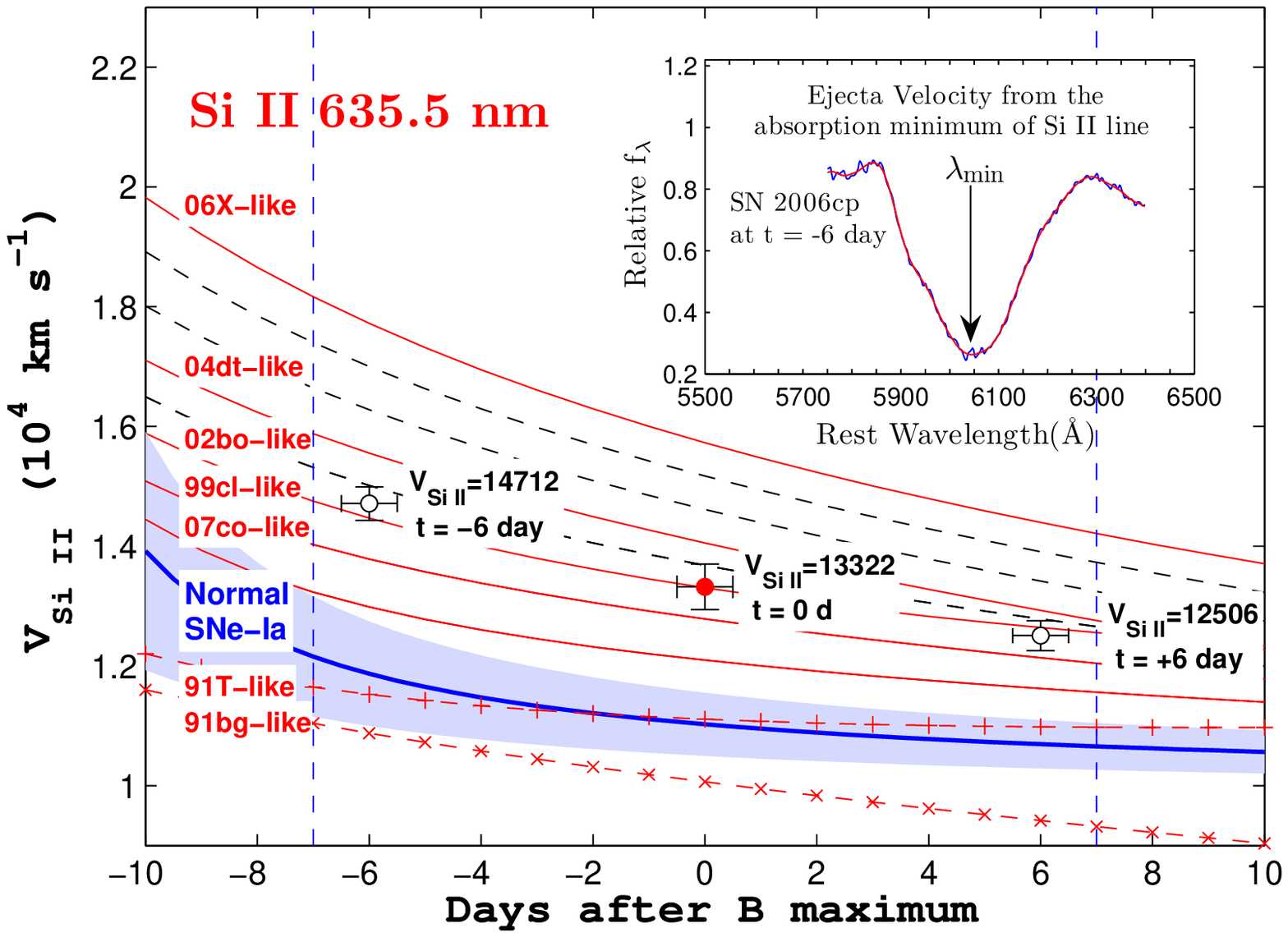} \\
\noindent {\bf Fig. S2:}
An example of the Si~II velocity measurement and normalization used in
this study. The absorption feature shown in the inset window is Si~II
635.5 nm from a spectrum of SN 2006cp at $t = -6$ day from maximum
light. The observed spectrum is the solid black curve, the locally
weighted scatterplot smoothed spectrum is the solid red curve, and the
absorption minimum is indicated by the position indicated by the down
arrow. The measured velocities (represented by the black filled dots)
are then normalized to the value at maximum light (indicated by the
red filled dot) by comparing with the template curves of the
well-observed SNe~Ia. The blue curve shows the mean evolution of NV
SNe Ia, and the gray region represents the $1\sigma$ uncertainty; the
red curves are part of the HV sample; the dashed-plus and dashed-cross
lines illustrate the evolution of the mean velocity for SN 1991T-like
and SN 1991bg-like events, respectively. These templates are
established with the published spectral data from the literature
({\it 2--4}). The black dashed curves represent the interpolated HV
templates from the neighboring observed ones, with an assumption of
proportional change.
\end{figure}

Following the classification scheme proposed by Wang et al. ({\it 19}), we divide the 188 SNe Ia into five groups: high velocity (HV), normal velocity (NV), overluminous SN 1991T-like (91T), subluminous SN 1991bg-like (91bg), and the extremely peculiar SN 2002cx-like (02cx) SNe Ia, with respective relative fractions of 21.2\%, 44.1\%, 6.9\%, 16.0\%, and 3.7\%. (8.1\% of SNe Ia could not be classified due to the absence of early-time spectra.) The HV:NV ratio, derived with the "season-optimal" subsample, may suffer some effects from the excluded sample. However, these effects cannot be directly addressed because we do not have early-time spectra of most of the excluded SNe and we were thus not able to subclassify them. The spectral data used for the subclassifications in this study are based primarily on the datasets published by the Center for Astrophysics Supernova Program ({\it 31, 33}) and the Berkeley Supernova Program ({\it 32}). For some objects the spectral parameters are taken from the IAU Circulars.

Table S1 lists the observed parameters as well as the classifications of the 188 SNe Ia. The meaning of each column is as follows: Column (1), the SN name; Column (2), the host-galaxy name; Column (3), the recession velocity (v) in the reference frame of the cosmic microwave background (CMB) when v $>$ 3000 km s$^{-1}$, or for a self-consistent Virgo-centric infall vector of 220 km s$^{-1}$ when v $<$ 3,000 km s$^{-1}$, using the procedure given at the NASA Extragalactic Database (NED); Column (4), the host-galaxy numerical Hubble type, or "T" type ({\it 54}); Column (5), the host-galaxy absolute $K$-band magnitude taken from The 2MASS Redshift Survey (55), dereddened by the Galactic reddening using the full-sky maps of dust infrared emission (38) and the Cardelli et al. reddening law with R$_{V}$ = 3.1 (56); Column (6), the host-galaxy physical radius at the 25.0 $B$-mag arcsec$^{-2}$ isophote; Column (7), the deprojected galactocentric distance of the SN in its host galaxy, in units of the galaxy radius R$_{gal}$; Column (8), the velocity of Si II line measured around maximum light, and its uncertainty; Columns (9)-(11), the u'$-$, g'$-$, and r'$-$band fractional fluxes, measured with the SDSS ({\it 36}) images ({\it 57}), representing the sum of counts registered in all pixels with fewer counts than measured at the SN location divided by the sum of all counts associated with the galaxy ({\it 35, 58}); and Column (12), the SN Ia subtype.

\vskip 1.0 cm
{\bf S2. Some Selection Effects in the Determination of the Radial Distance}
\vskip 0.5 cm
In calculating the SN relative distance, we assumed that galaxies are circular disks and that they exhibit different major and minor axes only because of different inclinations. This assumption can be used to derive the deprojected radial distance of the SNe in spiral galaxies with an intermediate inclination angle (e.g., $<$ 70$^{\circ}$. For galaxies with large inclination angles, such a tilt correction may cause large uncertainties in an attempt to derive the deprojected distance. As an alternative, we can also make an assumption of spheroids for all of the galaxies. Based on the projected radial distance, we reanalyzed the data and found that the HV SN Ia sample is still highly concentrated near the galaxy center relative to the NV sample, with a K-S probability of 0.3\% that they come from the same parent population.

\begin{figure}
\includegraphics[width=16.0cm]{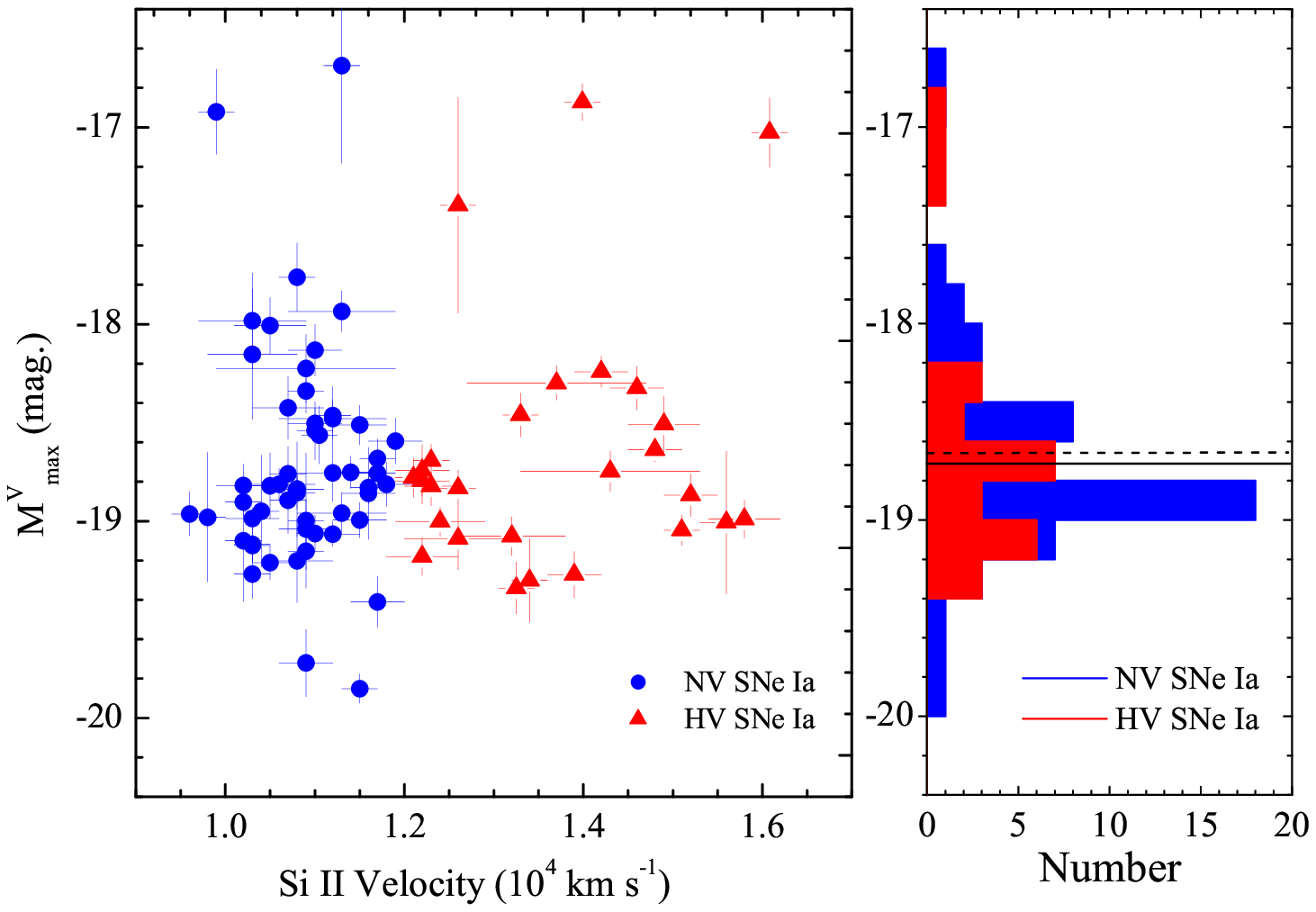} \\
\noindent {\bf Fig. S3:}
Left panel:  The V-band absolute magnitudes, corrected for Galactic reddening, 
plotted as a function of the Si II velocity around maximum light. 
Right panel: Histograms of $V$-band absolute magnitudes for the HV and NV subclasses of 
SNe Ia. Two vertical lines in the panel mark the respective mean values of the absolute 
magnitudes for the NV (solid) and HV (dashed) SNe Ia, which are  18.71$\pm$0.08 mag and 
18.64$\pm$0.12 mag. A K-S test rejects the hypothesis with a probability of 99.6\% that 
these two SN Ia populations have different peak luminosities.
\end{figure}

We further considered possible selection effects that may bias the radial distribution of the HV and NV SNe Ia, which can arise from differences in SN luminosity and size of the host galaxy. For example, brighter SNe are more easily detectable in the central regions of galaxies than fainter SNe. In Figure S3, we compared the distribution of the V-band magnitudes M$^{V}_{max}$, corrected for Galactic reddening, for the HV and NV SN Ia samples. The mean value and the distribution of M$^{V}_{max}$ are found to be comparable for these two samples. This indicates that the selection effect due to the different luminosity distributions of HV and NV SNe Ia should not have a significant bias on the radial distribution in their host galaxies.

According to the analysis of the LOSS SN sample ({\it 23}), there is a higher probability that the SNe tend to be missed in the central regions of smaller galaxies relative to that of larger galaxies. We further examined this possible selection effect by excluding the smaller galaxies from our analysis. Restricting the sample used in the analysis to the SNe Ia with major axis of the host galaxy greater than 1.0' (a total of 96) and 1.3' (a total of 56), we found that the differences in radial distributions between the HV and NV subsamples are still statistically significant, with K-S probabilities of 0.4\% and 5.7\% that they come from the same population.

The above analysis indicates that the concentration of HV SNe Ia in the central regions of galaxies cannot be explained with any known selection effects, so it seems to be a property intrinsic to these objects.

\vskip 1.0 cm
{\bf S3. Measurements of the Fractional Flux at the SN Location}
\vskip 0.5 cm

The fractional flux is measured in host-galaxy images as the fraction of total host light in pixels fainter than or equal to the light in the pixel at the location of the SN ({\it 35}). Were the SNe of different types to exactly track the host-galaxy light, their histograms would follow the diagonal line. This parameter is independent of galaxy morphology and provides a better way to quantify the correlation between SNe and the light of their hosts. The host-galaxy images in our analysis were taken from the SDSS DR 6 ({\it 56}), which includes 54-s exposures in the Sloan filter set. To avoid possible contamination by residual light from the SNe, we excluded those images in which the SDSS observation of the host was made during the period from 1 month before to 12 months after the SN detection. The SDSS u'g'r'$-$band images were used for the measurements; they are available for 25 HV SNe Ia, 39 NV SNe Ia, 39 SNe Ibc, and 102 SNe II.

Following Ref. (57), we identify the pixels with host-galaxy light by selecting contiguous pixels having signal-to-noise ratios greater than 1. Before that, the foreground stars were masked out and the enclosed pixel values were replaced with the mean of pixel values around the mask perimeter. From the distribution of pixel values of the galaxy light in the SDSS u'g'r' bands and the corresponding values at the SN location, we calculate the fractional fluxes. All of the results are listed in columns (10)-(12) of Table S1. Note that the u'-band fractional flux may suffer larger uncertainties because the SDSS images are not deep enough in this band to allow accurate measurements.

\vskip 1.0 cm
{\bf S4. Redshift and Morphology of the Host Galaxies}
\vskip 0.5 cm
\begin{figure}
\includegraphics[width=16.0cm]{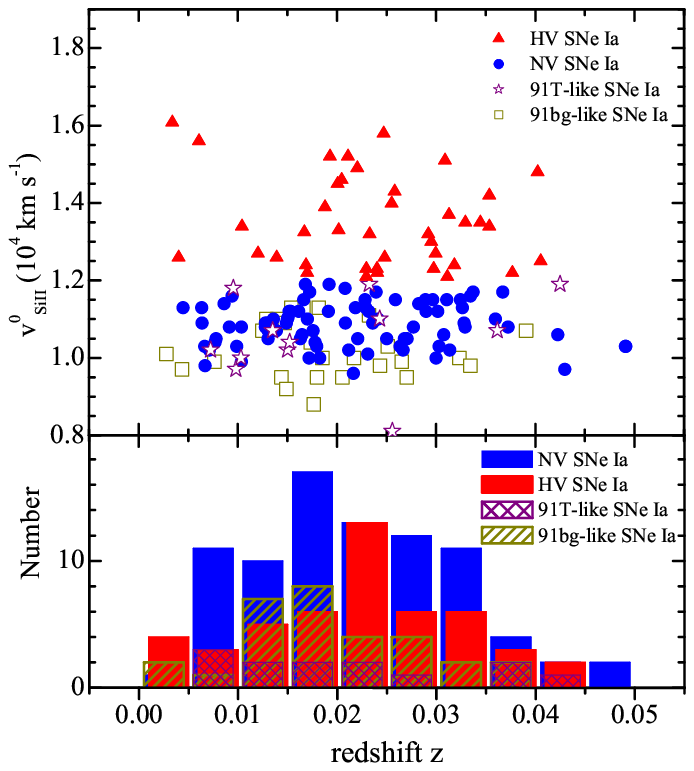} \\
\noindent {\bf Fig. S4:}
A plot of the Si II velocities of SNe Ia near maximum light and the 
redshift of their host galaxies. In the upper panel, the HV SN Ia hosts 
are represented as red triangles, the NV SN Ia hosts as blue dots, the 
91T-like hosts as purple stars, and the 91bg-like hosts as dark-yellow squares. 
The lower panel shows a histogram for the redshift distribution of the hosts of 
different types. The mean redshift for the HV SN Ia sample is 0.024, comparable 
to that of the NV sample (mean value 0.022)..
\end{figure}

\begin{figure}
\includegraphics[width=16.0cm]{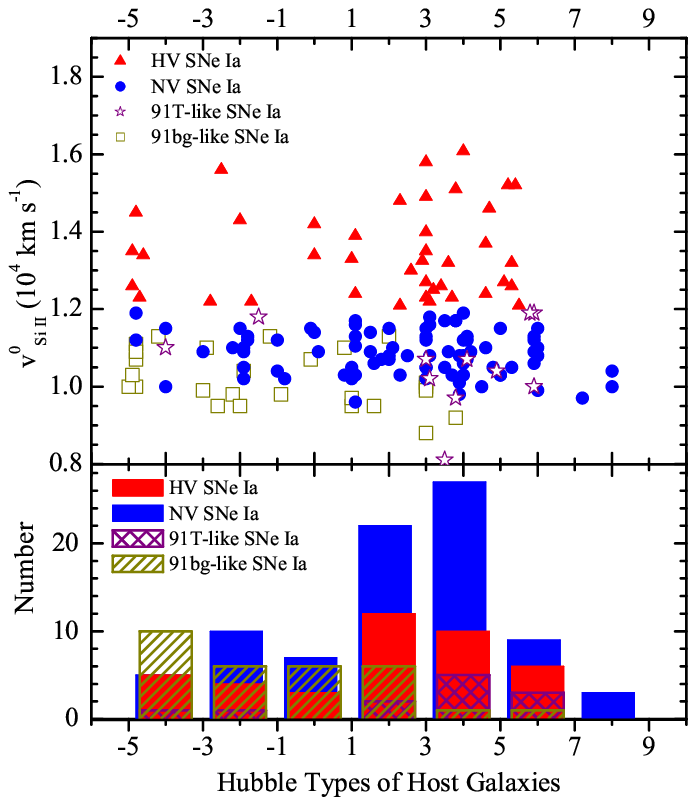} \\
\noindent {\bf Fig. S5:}
A plot of the Si II velocities of SNe Ia near maximum light and the
numerical Hubble types of their host galaxies.  The symbols are the
same as in Fig. S4. The plot is then projected onto the lower panel,
where a histogram distribution is displayed for each Hubble type of the
host populations. The K-S test gives a probability of 70\% that the
HV hosts and NV hosts have the same morphological distribution.

\end{figure}

Figure S4 shows the redshift distribution of the host galaxies. The mean redshift distribution is estimated as z = 0.0240 for the HV SNe Ia and 0.0217 for the NV SNe Ia; the HV SN Ia hosts are, on overage, at slightly larger distances. Such a small difference cannot account for the paucity of the HV SNe Ia in smaller galaxies, as the mean absolute $B$-band magnitude and the major axis of the host galaxies are found to increase by only $\sim$0.08 mag and $\sim$0.8 kpc from z = 0.0217 to 0.0240. The Hubble type ("T" type) of the host galaxies of the HV and NV SN Ia samples is also compared in Figure S5. A K-S test gives a probability of 70\% that these two samples have a similar morphology distribution, indicating that the differences in light radii and luminosity of galaxies seen in Figure 4 are not due to an observational bias.

\begin{figure}
\includegraphics[width=16.0cm]{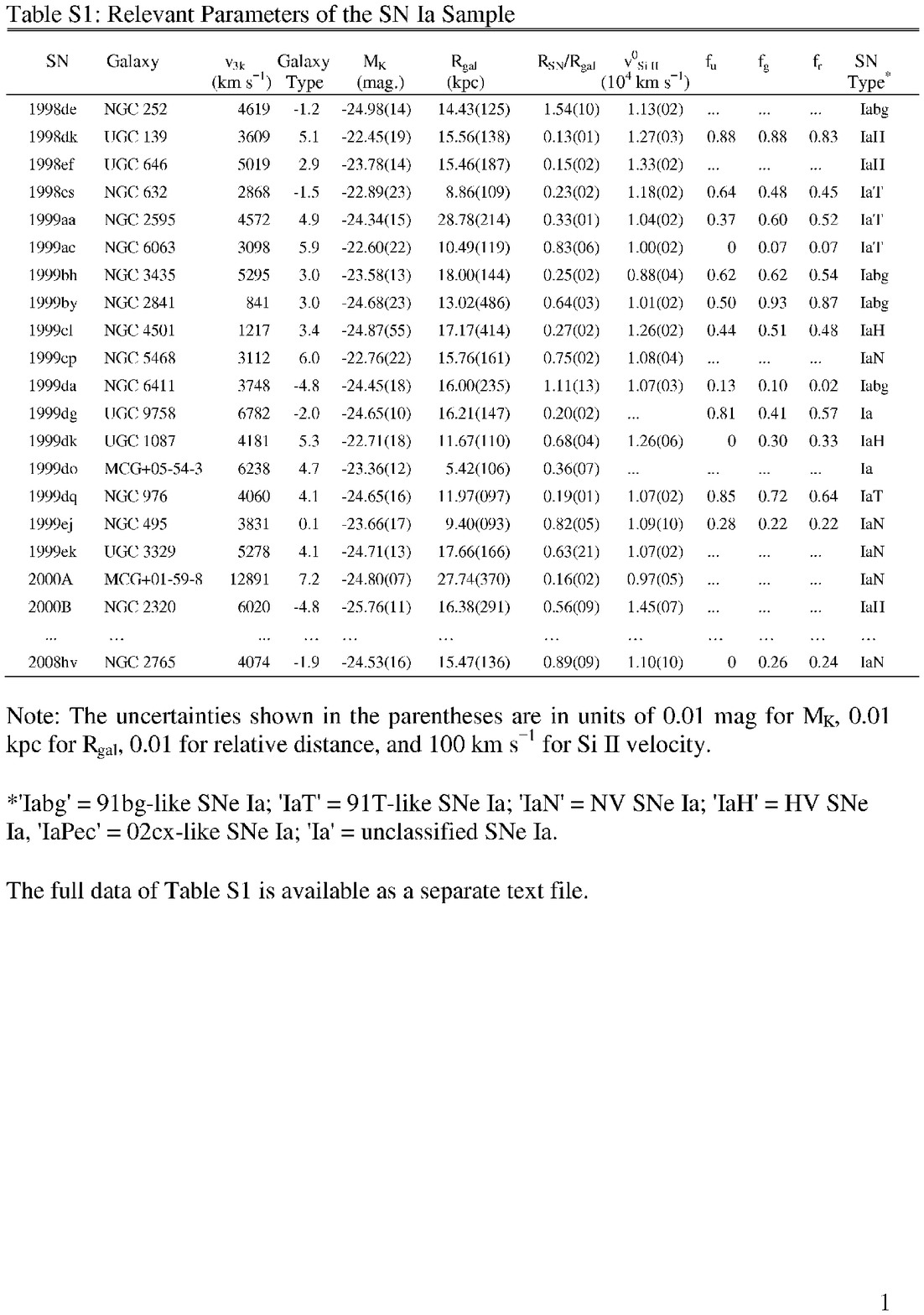} \\

\end{figure}

\subsection*{References}
\begin{itemize}
\item[52.] W. D. Li, et al. {\it Mon. Not. R. Astron. Soc.} {\bf 412}, 1473 (2011).
\item[53.] W. S. Cleveland, {\it J. Am. Statist.} {\bf 74}, 368 (1979).
\item[54.] G. de Vaucouleurs, A. de Vaucouleurs, H. G. Corwin, R. J., Buta, G. Paturel, and P. Fouque, Third Reference Catalogue of Bright Galaxies (New York: Springer-Verlag) (1991).
\item[55.] J. P. Huchra, et al. {\it Astrophys. J. Suppl. Ser.} {\bf 199}, 26 (2012).
\item[56.] J. A. Cardelli, G. C. Clayton, \& J. S. Mathis, {\it Astrophys. J.} {\bf 345}, 245 (1989).
\item[57.] J. K. Adelman-McCarthy, et al.  {\it Astrophys. J. Suppl. Ser.} {\bf 175}, 297 (2000).
\item[58.] P. L. Kelly, R. P. Kirshner, \& M. Pahre, {\it Astrophys. J.} {\bf 687}, 1201 (2008).
\end{itemize}

\end{document}